\documentstyle[eqsecnum,prl,aps,epsf]{revtex}
\begin{document}

\draft

\title{Projected $SO(5)$ Models}
\author{Shou-Cheng Zhang, Jiang-Ping Hu}
\address{
Department of Physics,
Stanford University,
Stanford, CA 94305
}
\author{Enrico Arrigoni, Werner Hanke}
\address{Institut f\"{u}r Theoretische Physik, 
Universit\"at W\"{u}rzburg, Am Hubland,\\
D-97074 W\"{u}rzburg, Federal Republic of Germany}
\author{Assa Auerbach}
\address{Department of Physics, Technion, Haifa 32000, Israel}
\date{\today}
\maketitle
\begin{abstract}
We construct a class of projected $SO(5)$ models where the 
Gutzwiller constraint of no-double-occupancy is implemented 
exactly. We introduce the concept of projected $SO(5)$ symmetry
where all static correlation functions are exactly $SO(5)$ 
symmetric and discuss the signature of the projected $SO(5)$
symmetry in dynamical correlation functions. 
We show that this class of projected
$SO(5)$ models can give a realistic description of the global phase
diagram of the high $T_c$ superconductors and account for many of their
physical properties.  
\end{abstract}


\section{Introduction}
\label{sec:I}

Recently, a unified theory of antiferromagnetism (AF) and 
superconductivity (SC) has been proposed for the high $T_c$
cuprates\cite{so5}. This theory is based on the $SO(5)$ symmetry
between AF and SC, and offers a unified description of the 
global phase diagram for this class of materials. 
While the theory was originally proposed
as a effective field theory description, it was soon realized
that the $SO(5)$ symmetry could be implemented exactly
at a microscopic level\cite{henley,rabello,burgess2,szh,eder2}, 
and it can also be checked numerically 
in common strongly correlated models such as the $t-J$ 
model\cite{meixner,eder,hanke-review}.
While the phase diagram\cite{so5,assa,burgess1,hu_xiao} and collective 
excitations\cite{resonance1,resonance2} in the
SC state derived from these $SO(5)$ models bear strong resemblance 
with the high $T_c$ cuprates, and a number of novel experimental
predictions have been made\cite{junction1,vortex,junction2,goldbart1,goldbart2}, 
the Mott insulating behavior at
half-filling is a puzzling aspect which challenges the fundamental
validity of the $SO(5)$ models\cite{greiter,greiter-reply,anderson,sns}. 
To be more precise, the exact 
$SO(5)$ symmetry requires collective charge two excitation at
half-filling to have the same mass as the collective spin wave
excitations. This condition is clearly violated in a Mott insulating
system where all charge excitations measured with respect to a particle
hole symmetric point have a large energy gap of few $eV$, while 
the spin wave excitations are massless. In the original $SO(5)$ proposal,
it was pointed out that this situation is analogous to a easy-axis 
antiferromagnet in a external uniform field, and a $SO(5)$ 
symmetry breaking term at half-filling
was introduced in order to describe this 
asymmetric behavior between spin and charge. The
chemical potential also introduces a $SO(5)$ symmetry breaking
term, however, it was shown that these two terms could
compensate each other\cite{so5,eder} so that the {\it static potential} governing
the $SO(5)$ superspin could still be $SO(5)$ symmetric.

Since the asymmetry between the charge and spin excitations at half-filling
is of the order of the Coulomb energy scale $U$, the $SO(5)$ symmetry
breaking terms must also be of that order. Since there are various types
of symmetry breaking terms, one might hope that their effects could
partially cancel each other to arrive at a qualitatively
correct picture. However, this type of cancellation is very delicate,
and approximate calculations could easily lead to erroneous 
conclusions. In particular, one is interested in which physical 
properties could exhibit $SO(5)$ symmetric properties in the limit
when the Coulomb gap is taken to infinity. For example, one could
ask the following questions:

1) One of the hallmarks of the $SO(5)$ symmetry is not only the degeneracy
between the AF and SC states at a given chemical potential, but the
approximate degeneracy among all mix-states interpolating between AF and SC,
{\it i.e.} the independence of the ground state energy on the superspin
angle. What is the potential barrier separating the AF and SC states at
their degeneracy point in the limit $U\rightarrow \infty$? If there is 
a large energy barrier in this limit, one would argue that the concept
of $SO(5)$ symmetry is not a useful one, at least not for quantitative
calculations. On the other hand, if the potential barrier is finite and
small in the $U\rightarrow \infty$ limit, the concept of a approximate
$SO(5)$ symmetry would be a useful one. 
  
2) Exact $SO(5)$ symmetry predicts four massless collective modes. In the
half-filled AF state, besides the two conventional massless spin wave 
modes, the exact $SO(5)$ symmetry predicts a massless 
doublet of $\pi^\pm$ modes, with charge $\pm 2$. However, a Mott 
insulator has a large gap to all charge excitations. Therefore, it is
clear that one of the $\pi^\pm$ has to be projected out of the spectrum
in the limit $U\rightarrow \infty$, say the $\pi^+$ mode carrying
charge $+2$. What happens to the rest of the Goldstone modes, the 
$\pi^-$ mode carrying charge $-2$ and the $\pi^\alpha$ triplet mode of the 
SC state? In the $U\rightarrow \infty$ limit, can they all be 
simultaneously massless at the transition point between AF and SC?
Since the pure SC state can only be reached with a finite doping 
concentration, is it possible that the Gutzwiller projection does not
affect the $\pi^\alpha$ triplet mode of the SC state? 

In order to address these questions, it is desirable to construct a 
low energy effective theory without any parameters of the order of
the Coulomb scale $U$. 
In this work, we construct a class of projected $SO(5)$ models which treat
the Gutzwiller constraint exactly and locally on every site. 
We use this model to answer
the physical questions posed above and show that the answers are affirmative.
In the $U\rightarrow \infty$ limit, when the Gutzwiller constraint is 
implemented exactly, the ground state energy can still be $SO(5)$ symmetric
and independent of the superspin direction. After projecting out the $\pi^+$
mode, all other Goldstone modes remain massless at the symmetric point. 
The dispersion relation of the collective modes bear unique 
signature of the projected $SO(5)$ symmetry. Furthermore, the
$\pi^\alpha$ triplet modes of the pure SC states are unaffected 
by the Gutzwiller projection. These
properties define the concept of a projected $SO(5)$ symmetry ($pSO(5)$),
whose properties and consequences we shall explore in this paper.

The fundamental quantity in the $SO(5)$ theory is the locally
defined five component superspin vector $n_a(x)=(n_1,n_2,n_3,n_4,n_5)$ 
describing the local AF and SC order parameters respectively. 
In the nonlinear $\sigma$ model formalism, these are treated 
as mutually commuting coordinates
and their dynamics is given by their conjugate 
momenta $p_a(x)=(p_1,p_2,p_3,p_4,p_5)$. The
charge operator is the angular momentum in the $n_1-n_5$ plane:
\begin{equation}
Q (x) = L_{15} = n_1 p_5 - n_5 p_1
\label{charge}
\end{equation}
Implementing the Gutzwiller constraint corresponds to requiring
\begin{equation}
Q(x) \le 0
\label{Gutzwiller}
\end{equation}
for every local $SO(5)$ rotor. From equations
(\ref{charge}) and (\ref{Gutzwiller}) and subsequent discussions, 
we shall see that the Gutzwiller projection in the $SO(5)$ formalism
corresponds to going
from a fully symmetric $SO(5)$ rotor model to a {\it chiral}
$SO(5)$ rotor model, where both the static potential of the individual
rotors and the coupling between the rotors are still $SO(5)$ 
symmetric, but the rotors are constrained to rotate only in one
sense in the $n_1-n_5$ plane, consistent with (\ref{Gutzwiller}). 
This observation reveals a deep connection
between the Gutzwiller projection and the Lowest-Landau-Level (LLL)
projection in the fractional quantum Hall effect\cite{qhe-book}. 
To be more
precise, the Gutzwiller projection represented by equations
(\ref{charge}) and (\ref{Gutzwiller}) is 
analogous to the LLL projection, where all states in the LLL
have a definite sign of angular momentum. The LLL projection
can be analytically implemented by separating the 
cyclotron degrees of freedom
from the guiding center degrees of freedom, which amounts to
changing the commuting property between the $X$ and $Y$ coordinates
to a canonically conjugate commutation relation:
\begin{equation}
[X,Y] = i \l_0 
\label{guiding-center}
\end{equation}
where $l_0$ is the Landau length.
Exploiting this analogy, we find that the original $SO(5)$ model can be
fully Gutzwiller projected without changing its form, if one imposes 
the simple quantization condition between the superconducting components
of the superspin vector:
\begin{equation}
[n_1,n_5] = i/2  
\label{non-commute}
\end{equation}
In the symmetric $SO(5)$ model, the wave function of the $SO(5)$ rotors
are functions of the local coordinates $n_1$ and $n_5$, while the 
projected $SO(5)$ model only depends on their holomorphic combination
$z=n_1-i n_5$ and is independent of their anti-holomorphic 
combination $\bar z=n_1+i n_5$. 
This way, we arrive at a natural projection of the $SO(5)$ model 
where the local Gutzwiller constraint is taken into account exactly,
and the resulting model is free of the large Coulomb $U$ parameter.
Because the functional form of the symmetric $SO(5)$ model remain
the same and only the quantization condition is modified upon projection,
many important properties associated with the $SO(5)$ symmetry remain.
The central hypothesis of the $SO(5)$ theory is that this projected 
model is quantitatively accurate in describing both the static and dynamic
properties of the high $T_c$ cuprates, and we shall compare the properties
of this model with the phenomenology of the high $T_c$ systems.

\section{Construction of projected $SO(5)$ Models}
\label{sec:II}

\begin{figure*}[h]
\centerline{\epsfysize=4.0cm
\epsfbox{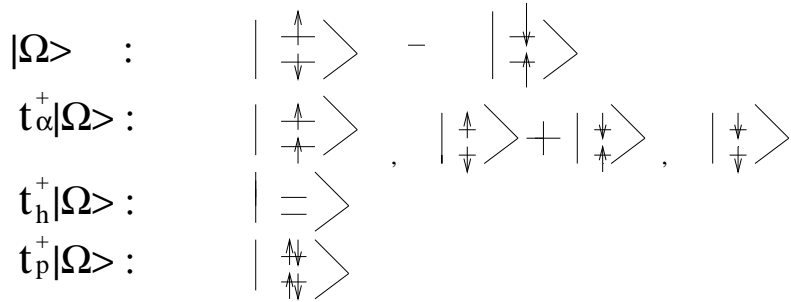}
}
\caption{Schematic representation of the singlet state, the triplet
magnon states and the hole and particle pair states.}
\label{fig1}
\end{figure*}

We begin with the symmetric $SO(5)$ Hamiltonian defined on a lattice,
\begin{eqnarray}
H &=& 
\Delta \sum_{ x } L_{ab}^2(x) 
- J \sum_{ <xx'> } n_a(x) n_a(x')
\nonumber \\
&+&  V \sum_{ <xx'> }
L_{ab}(x) L_{ab}(x')
\label{sigma}
\end{eqnarray}
where $n_a(x)$ denotes the five component superspin vector on a 
given site, and $L_{ab}(x)$ is the $SO(5)$ symmetry generator,
\begin{equation}
L_{ab} = n_a p_b - n_b p_a
\end{equation}
expressed here in terms of the superspin vector $n_a$ and its 
canonically conjugate momenta $p_a$,
\begin{equation}
[n_a, p_b] = i \delta_{ab}
\end{equation}
This lattice quantum non-linear $\sigma$ model can be rigorously derived
as the low energy limit of a microscopic $SO(5)$ ladder model\cite{szh,eder2}.
On the ladder,  the rung $SO(5)$ singlet state is the vacuum
$|\Omega\rangle$, from  which  the lowest SO(5) multiplet
$| a\rangle=   t_a^\dagger|\Omega\rangle$ is created by a quintet
of Bose creation operators, which satisfy

\begin{equation}
[t_a, t_b^\dagger] = \delta_{ab},~~~~~ t_a |\Omega\rangle=0
\end{equation}
Here, $a=2,3,4$
denote the triplet (magnon) states, and  $a=1,5$ are the hole and particle pair states.
(see Fig. (\ref{fig1})).
The superspin coordinates are microscopically constructed using  these lattice bosons,
\begin{equation}
n_a = \frac{1}{\sqrt{2}} (t_a + t_a^\dagger)  \ \ \
p_a = \frac{1}{i\sqrt{2}} (t_a - t_a^\dagger)
\end{equation}

Due to their microscopic origin, these bosonic states 
are hard-core bosons, in the sense that one can not define two of them on the
same rung. The $\Delta$ term in equation (\ref{sigma}) 
describes the gap energy of the
magnon and the pair states, the $J$ term stands for the hopping and the
spontaneous creation/destruction process of these states, and the $V$
term describes their nearest-neighbor interaction.  
This quantum nonlinear $\sigma$ model can in principle 
also be derived in higher dimensions from a microscopic $SO(5)$ 
symmetric model\cite{henley,rabello,burgess2}, by 
introducing a superspin vector as a
Hubbard-Stratonovich decoupling field, and integrate out the fermionic 
degrees of freedom in a gradient expansion. However, we shall proceed 
more heuristically here\cite{eder3}. For a two dimensional system, one can
imagine that the quantum $\sigma$ model Hamiltonian is obtained from a ``block spin"
type of coarse-graining of the microscopic electron Hamiltonian, and is defined 
on a lattice with twice the lattice constant compared to the microscopic electron
model. (This doubled unit cell is the minimal size needed to define the
local AF and d wave SC order parameters).
Therefore, each site $x$ in the effective model correspond
to a plaquette of
the microscopic electron model. (On a ladder, this corresponds to going from the
lattice sites to ladder rungs). The $SO(5)$ singlet state $|\Omega\rangle$ corresponds
to a ``RVB" type of singlet state, while the five-fold states $t_a^\dagger|\Omega\rangle$
describe the triplet magnon states, and the $d$-wave hole and particle pair states
on a plaquette. Unlike the ladder case, the magnon and the $d$- wave pair states
could condense in the ground state to form AF and SC broken symmetry states.
In fact, Eder\cite{eder3} has recently shown that properties of the AF states
can be described by a coherent state of magnon condensation on top of a uniform spin
liquid state. Our model therefore describes competition among the ``RVB" type 
of singlet vacuum and the two forms of broken symmetry order.

While it is reasonable to take $J$ and $V$ to be approximately equal for magnons and
pairs, the gap energy $\Delta$ for the neutral magnons and the charged pairs
are very different in the insulating state at half-filling. In fact, their difference
is of the order of the insulating gap $U$ at half-filling. 
Taking into account the hard-core condition and neglecting the nearest-neighbor interaction
$V$ for now (it has higher powers of time and space derivatives in
the continuum limit), we can express the general 
anisotropic $SO(5)$ model as
\begin{eqnarray}
H &=& 
\Delta_s \sum_{ x } t_\alpha^\dagger t_\alpha(x) + 
\Delta_c \sum_{ x } t_i^\dagger t_i(x) \nonumber \\
&-& J_s \sum_{ <xx'> } n_\alpha(x) n_\alpha(x') -
J_c \sum_{ <xx'> } n_i(x) n_i(x')
\label{anisotropic}
\end{eqnarray}
In this paper we shall use the convention where $a,b,..=1,2,3,4,5$ denote
the superspin indices, $\alpha,\beta,..=2,3,4$ denote the spin indices
and $i,j=1,5$ denote the charge indices, and repeated indices are summed over.
The main focus of our paper is to consider the limit where 
$\Delta_c >> \Delta _s$.

Let us define the charge eigen-operators $t_h$ and $t_p$ as
\begin{equation}
t_1 = \frac{1}{\sqrt{2}} (t_h + t_p)  \ \ \ 
t_5 = \frac{1}{i\sqrt{2}}(t_h - t_p) 
\label{p-h}
\end{equation}
 From this definition, it is clear that $t_h^\dagger$ is the 
creation operator for a hole pair and $t_p^\dagger$ is the 
creation operator for a particle pair. We can introduce a 
chemical potential term 
\begin{equation}
H_\mu = \mu \sum_x (t_p^\dagger t_p(x) - t_h^\dagger t_h(x))
\end{equation}
to describe the effects of doping. In the presence of this
chemical potential term, the gap energy of the hole
and particle pairs are $\Delta_c - \mu$ and $\Delta_c + \mu$
respectively. A chemical potential of the order of the charge
gap $\Delta_c$ is needed to induce a metal-insulator transition
in this system. Near such a transition point, the gap
energy of the hole pair 
\begin{equation}
\tilde \Delta_c = \Delta_c - \mu
\end{equation}
can be comparable to the spin gap $\Delta_s$, while the
gap towards a particle pair excitation is of the order of twice
the charge gap, and needs to be projected out of the spectrum
in the low energy limit. 

Therefore, within this formalism, the Gutzwiller projection is
equivalent to restricting ourselves to the projected Hilbert 
space where
\begin{equation}
t_p(x) |\Psi\rangle = 0
\label{constraint1}
\end{equation}
at every site $x$. Within this projected Hilbert space, the
projected Hamiltonian takes the form
\begin{eqnarray}
H &=& 
\Delta_s \sum_{ x } t_\alpha^\dagger t_\alpha(x) + 
\tilde \Delta_c \sum_{ x } n_i(x) n_i(x)  \nonumber \\
&-& J_s \sum_{ <xx'> } n_\alpha(x) n_\alpha(x') -
J_c \sum_{ <xx'> } n_i(x) n_i(x')
\label{Hamiltonian}
\end{eqnarray}
This Hamiltonian has no parameters of the order of $U$, and it is
reasonable to expect $\Delta_s\sim\tilde\Delta_c$ and $J_s\sim J_c$.
We see that the form of the Hamiltonian hardly changes from the
unprojected model, but the
definition of $n_1$ and $n_5$ is changed from 
\begin{eqnarray}
n_1 &=& \frac{1}{\sqrt{2}} (t_1 + t_1^\dagger) = 
\frac{1}{2} (t_h + t_p + t_h^\dagger + t_p^\dagger) \nonumber \\
n_5 &=& \frac{1}{\sqrt{2}} (t_5 + t_5^\dagger) = 
\frac{1}{2i} (t_h - t_p - t_h^\dagger + t_p^\dagger)
\label{before}
\end{eqnarray}
to
\begin{equation}
n_1 = \frac{1}{2} (t_h + t_h^\dagger) \ \ \
n_5 = \frac{1}{2i} (t_h - t_h^\dagger)
\label{after}
\end{equation}
 From equation (\ref{before}), we see that $n_1$ and $n_5$ commute 
with each other before the projection. However, after the projection,
they acquire a nontrivial commutation relation, as can be seen from
equation (\ref{after}):
\begin{equation}
[n_1,n_5] = i/2 
\end{equation}
Therefore, the Gutzwiller projection can be analytically implemented
in the $SO(5)$ theory by retaining the form of the Hamiltonian and
change only the quantization condition.

\section{Analogy with lowest-Landau-level projection}
\label{LLL}
The discussions outlined above reveal a deep connection between
the Gutzwiller projection within the $SO(5)$ formalism and the 
projection onto the lowest Landau level (LLL) in the context of 
the fractional quantum Hall effect. Consider the 
problem of a charged particle
in a strong magnetic field $B$ and a rotationally symmetric potential
$V(X,Y)$. In the absence of a magnetic field, all eigenstates form
irreducible representations of the two dimensional rotation 
group $O(2)$, characterized by integral eigenvalues of the angular
momentum operator
\begin{equation}
L_Z = X P_Y - Y P_X
\end{equation}
However, in the presence of a strong magnetic field and projected into
the LLL, only negative eigenvalues of $L_Z$ are realized. This is 
analogous to the situation encountered here. The local charge operator
in the $SO(5)$ theory takes the form of the angular momentum in the
$n_1-n_5$ plane as given by equation (\ref{charge}). When doubly
occupied sites are locally projected out, the local charge operator, or
the angular momentum in the $n_1-n_5$ plane, takes only negative values.
Since the chemical potential couples directly the angular momentum in the 
$n_1-n_5$ plane, it plays the role of a fictitious magnetic field
threading every $SO(5)$ rotor in the $n_1-n_5$ plane. The Landau level
spacing $\hbar \omega_c$ is analogous to the charge gap $\Delta_c$
encountered here, and both are taken to be infinity in the projected
models. After the projection, the Hamiltonian in the Landau level problem
retains its $O(2)$ symmetric form,
\begin{equation}
H = V(X,Y)
\end{equation}
although a new quantization condition is imposed between $X$ and $Y$, as
given by equation (\ref{guiding-center}). This is analogous to the 
observation we made here that the Hamiltonian formally retains a
$SO(5)$ symmetric form after the projection (\ref{constraint1}), but the
quantum dynamics is changed due to the non-trivial commutator between
$n_1$ and $n_5$. In both cases only a part of the full symmetry multiplets
remain after the projection. However, the formal symmetry of the 
Hamiltonian has direct physical manifestations despite the projection.
For example, in the LLL problems, semi-classical orbits of the guiding
center coordinates are still $O(2)$ symmetric. In our case, we shall see
that the static potential for the superspin vector can still be $SO(5)$
invariant despite the projection.   
 
Perhaps the most explicit way to establish the precise connections between
these two problems is to consider the constraints on the wave function. 
In the symmetric gauge of the LLL problem, the annihilation operator for
the cyclotron coordinates takes the form\cite{qhe-book}
\begin{equation}
a= \partial_{\bar z} + z/4
\end{equation}
where $z=X+iY$ and $\bar z=X-iY$. Projection onto LLL requires
\begin{equation}
a \Psi(z,\bar z) = 0
\label{constraint2}
\end{equation}
which determines the form of the LLL wave function to be 
\begin{equation}
\Psi(z,\bar z) = f(z) e^{-z\bar z/4}
\end{equation}
where $f(z)$ is a holomorphic function of $z$ only. 
This holomorphic condition also places strong constraints
in many-body systems and led to the celebrated Laughlin's
wave function. Our 
no-double-occupancy constraint (\ref{constraint1}) is
analogous to the LLL constraint (\ref{constraint2}).
In fact from equations (\ref{p-h}) and (\ref{before}),
we obtain
\begin{equation}
t_p = \frac{1}{2} ( z + 2 \partial_{\bar z})
\end{equation}
where $z=n_1-in_5$ and $\bar z=n_1+in_5$. For a single 
unprojected $SO(5)$ rotor, the wave function 
$\Psi(n_a)$ is a function of the superspin
coordinates. However, the Gutzwiller projection 
(\ref{constraint1}) restricts the wave function to be
\begin{equation}
\Psi(n_1,n_2,n_3,n_4,n_5) = f(z=n_1-in_5,n_2,n_3,n_4) e^{-z\bar z/2}
\end{equation}
where $f(z,n_2,n_3,n_4)$ is a holomorphic function of $z$.
For a collection of $SO(5)$ rotors, the superspin coordinates are
themselves functions of the lattice sites $x$, and 
$\Psi[n_a(x)]$ is a functional is the superspin
coordinates at each site. For the projected $SO(5)$ models, this
functional is restricted to take the form
\begin{equation}
\Psi[n_a(x)] = f(z(x),n_\alpha(x)) \prod_{x} e^{-z\bar z(x)/2}
\label{functional}
\end{equation}
where $f(z(x),n_\alpha(x))$ is a holomorphic functional of 
$z(x) = n_1(x)-in_5(x)$.

The formal but precise analogy between the two types of projection
allows us to introduce the concept of a chiral $SO(5)$ rotor. 
This is a system of rotors with $SO(5)$ invariant potential and
coupling, however, the rotation within the $n_1-n_5$ plane is 
chiral, {\it i.e.} only one sense of the rotation is allowed. 
Such a system of chiral $SO(5)$ rotors is described by the
wave functional in equation (\ref{functional}).

\section{$SO(5)$ Symmetry of the ground state energy}
\label{ground-state}

Having discussed the general notions of the projected $SO(5)$ model,
we are now in a position to explore the phase diagram of this model.
As we commented earlier, the projected $SO(5)$ model describes the
competition and unification of the spin liquid, AF and the SC states.
In the original unprojected $SO(5)$ symmetric model, not only are the AF
and SC states degenerate in energy, but they are also degenerate with 
all the intermediate coexistence states.
This points out a route from AF to SC with no potential barrier,
and introduces the concept that the metal-insulator transition in the
high $T_c$ systems can be viewed as a smooth rotation of the $SO(5)$
superspin. One of the key questions to be answered in this work is 
what happens to the picture in the case of projected $SO(5)$ symmetry.

In anticipation of the competition of the states discussed above, we 
construct a class of variational wave functions in the coherent 
state representation:
\begin{equation}
|\Psi\rangle = \prod_x (\cos\theta(x)+\sin\theta(x)(m_\alpha(x)t^\dagger_\alpha(x)
+\Delta(x) t^\dagger_h(x))) |\Omega\rangle
\label{variational}
\end{equation}
Here $|\Omega(x)\rangle$ denotes a local singlet state defined by 
$t_\alpha(x)|\Omega(x)\rangle=t_h(x)|\Omega(x)\rangle=0$ and $|\Omega\rangle$
is a product state of these local singlets, $|\Omega\rangle=\prod_x|\Omega(x)\rangle$. 
$\theta(x)$ is a local variational parameter describing the competition
between long range order and quantum disorder. 
For $\theta(x)=0$ our variational
wave function describe a spin singlet ground state, while a non-zero
value of $\theta$ describes a coherent state formed by the
local singlet, the magnon or the hole pair state. 
This wave function is a generalization of the coherent state
description of a AF state in terms of a magnon condensate\cite{eder2,eder3}.
As we shall see from the following equation (\ref{superspin_vector}),
$sin2\theta$ stands for the
length of the $SO(5)$ superspin vector. $m_\alpha(x)$ and $\Delta(x)$
are general complex variational parameters describing the local 
amplitude for magnons and hole pairs. We notice that this wave function 
satisfies both the Gutzwiller constraint (\ref{constraint1}) and
the hard-core constraint for magnons and hole pairs exactly.
It is easy to see that
\begin{eqnarray}
\langle\Psi|n_\alpha(x)|\Psi\rangle &=& \frac{1}{\sqrt{2}} sin2\theta(x) 
Re(m_\alpha(x))\nonumber \\
\langle\Psi|n_1(x)|\Psi\rangle &=& \frac{1}{2} sin2\theta(x) 
Re(\Delta(x))\nonumber \\
\langle\Psi|n_5(x)|\Psi\rangle &=& \frac{1}{2} sin2\theta(x) Im(\Delta(x))
\label{superspin_vector}
\end{eqnarray}
where $Re$ and $Im$ denote the real and imaginary parts of a complex 
number. The coupling terms in the projected $SO(5)$ Hamiltonian depend only
on $n_\alpha(x)$, $n_1(x)$ and $n_5(x)$. Therefore, the coupling energy
depends only on the real part of $m_\alpha(x)$ while it depends on both the
real and imaginary parts of $\Delta(x)$. Therefore, for discussing the ground state
wave functions, we can assume without
loss of generality that $m_\alpha(x)$ is real and $\Delta(x)=m_1(x)+im_5(x)$.
The normalization condition $\langle\Psi|\Psi\rangle=1$ can be implemented by 
the constraint that
\begin{equation}
m_a^2(x) = (m_1^2 + m_5^2 + m_\alpha^2)(x) = 1 
\label{normalization}
\end{equation}
Therefore, we see that although we have completely projected out the
particle pair states, the local degrees of freedom can still be represented
by a vector on a five dimensional sphere. 

Uniform states are obtained by taking all parameters to be constant. For 
$\Delta=0$ and $sin2\theta\neq 0$, our wave function $|\Psi\rangle$ 
describes a pure AF state with the following properties:
\begin{eqnarray}
\langle\Psi|Q|\Psi\rangle &=& \langle\Psi|\sum_x t^\dagger_h t_h(x)|\Psi\rangle =0 \nonumber \\
\langle\Psi|N_\alpha|\Psi\rangle &=& \langle\Psi|\sum_x n_\alpha(x)|\Psi\rangle 
                      = N \frac{1}{\sqrt{2}} sin2\theta \ m_\alpha \nonumber \\
\langle\Psi|S_\alpha|\Psi\rangle &=& \langle\Psi|\sum_x i\epsilon^{\alpha\beta\gamma}
                               t^\dagger_\beta t_\gamma(x)|\Psi\rangle =0 \nonumber \\
\langle\Psi|S^2|\Psi\rangle &=&  \langle\Psi|(\sum_x S(x))^2|\Psi\rangle
                 = N 2 \sin^2\theta +N \sin^4\theta(1-m^4_{\alpha})
\label{AF-state}
\end{eqnarray}
where $N$ is the number of lattice sites. Equation (\ref{AF-state}) describes a
half-filled state with a macroscopic Neel magnetization,
and vanishing uniform magnetization. Furthermore, this state is composed as a
linear superposition of eigenstates with different values of the total spin, and
the fluctuation of the total spin scales like 
$\sqrt{\langle S^2\rangle} \propto \sqrt{N}$, just as one expects
from a standard Neel state.
   
On the other hand, for $m_\alpha=0$ and $sin2\theta\neq 0$, $|\Psi\rangle$ 
describes a pure SC state with the following properties:
\begin{eqnarray}
& & \langle\Psi|Q|\Psi\rangle = \langle\Psi|\sum_x t^\dagger_h t_h(x)|\Psi\rangle =
           N \sin^2 \theta \nonumber \\
& & \langle\Psi|N_1 + i N_5|\Psi\rangle = \langle\Psi|\sum_x (n_1 + i n_5)(x)|\Psi\rangle 
                      = N \frac{1}{2} sin2\theta \ (m_1+im_5) \nonumber \\
& & \langle\Psi|Q^2|\Psi\rangle - \langle\Psi|Q|\Psi\rangle^2  = N \sin^2\theta \cos^2 \theta
\label{SC-state}
\end{eqnarray}
Equation (\ref{SC-state}) describes a state with a finite doping density
and a finite SC order parameter. Just as in the standard BCS case, 
this state is composed as a
linear superposition of eigenstates with different values of the total charge, and
the fluctuation of the total charge scales like 
$\sqrt{\langle Q^2\rangle-\langle Q\rangle^2} \propto \sqrt{N}$, 
just as one expects from a standard SC state.

However, besides these two {\it pure} states, there is a class of {\it mixed} states
which interpolates between the pure AF and SC states. Taking $m_1=\sin\alpha$ and
$m_2=\cos\alpha$, we see that the mixed states
have the following property:
\begin{eqnarray}
& & \langle\Psi|Q|\Psi\rangle = N \sin^2\theta \sin^2\alpha \nonumber \\
& & \langle\Psi|N_1 + i N_5|\Psi\rangle = N \frac{1}{2} sin2\theta \sin\alpha \nonumber \\
& & \langle\Psi|N_2|\Psi\rangle = N \frac{1}{\sqrt{2}} sin2\theta \cos\alpha
\label{mixed}
\end{eqnarray}
Therefore, we see that there is a continuous family of intermediate mixed states
interpolating between the pure AF state at half-filling and the pure SC
state with finite doping density. As the $SO(5)$ angle $\alpha$ rotates continuously
from a pure AF state with $\alpha=0$ to a pure SC state with $\alpha=\pi/2$, 
the hole density of the mixed state interpolates 
continuously between these two limits. Therefore, our wave function gives a unified
description of AF and SC and points out a precise route from AF to SC as 
the doping level is varied. In order for this route, or small deviations from this
route, to be physically realized in the high $T_c$ superconductors, we have to 
demonstrate that there is no large energy barrier for the intermediate mixed states,
or that the ground state energy is approximately independent of the $SO(5)$
mixing angle $\alpha$. In particular, we have to show that the energy barrier is
independent of the Hubbard energy $U$, in the limit of large $U$. In the following, we
shall investigate this question.

The energy functional $\langle\Psi|H|\Psi\rangle$
describes the coupling between these five dimensional vectors $m_\alpha(x)$, and
it is given by 
\begin{eqnarray}
& & \langle\Psi|H|\Psi\rangle = E(\theta(x),m_a(x)) \nonumber \\
&=& -\frac{J_s}{2} \sum_{xx'} sin2\theta(x) sin2\theta(x') m_\alpha(x) m_\alpha(x')
- \frac{J_c}{4}\sum_{xx'} sin2\theta(x) sin2\theta(x') m_i(x) m_i(x') \nonumber \\
& & + \Delta_s \sum_x \sin^2\theta(x) m_\alpha^2(x)
+ \tilde \Delta_c \sum_x \sin^2\theta(x) m_i^2(x)
\label{energy}
\end{eqnarray}
This ground state energy functional describes a systems of coupled rotors
satisfying the constraint (\ref{normalization}). At the point $J_c=2J_s$ 
and $\tilde \Delta_c=\Delta_s$ in parameter space,
this rotor model is exactly $SO(5)$ symmetric. This is a central observation
of this work.
 
 From this consideration we learn a very important lesson about the compatibility
of the Mott insulating gap and the idea of a smooth $SO(5)$ rotation from AF to SC. 
As we have seen, the large asymmetry between the charge and spin gap at half-filling
necessitates the removal of the $Q(x)>0$ part of the $SO(5)$ multiplets, therefore,
the dynamics close to half-filling has to be modified. But the static potential 
governing the transition from AF to SC can remain $SO(5)$ symmetric, and in particular,
the energy barrier separating these two states can remain small in the limit where the
Mott insulating gap tends to infinity. 

We also observe a crucial difference between
the projected and the unprojected $SO(5)$ models. In the unprojected $SO(5)$ model
with the full $SO(5)$ symmetry, AF and SC states are degenerate at half-filling,
and the rotation between these two states can be continuously performed without changing
the density to going away from half-filling. This case is similar to the well-known 
degeneracy between the CDW state and the $s$ wave SC state for the negative $U$
Hubbard model at half-filling. In the projected $SO(5)$ model, where all particle
pair states have been locally removed, the $SO(5)$ rotation from the AF to SC states
are accompanied by the continuous change of the hole density, and a pure SC state
can only be reached at a finite critical hole doping density $\rho_c=\sin^2\theta$. 
While the unprojected $SO(5)$ symmetry is only valid at half-filling, the 
projected $SO(5)$ symmetry can be valid for a range of doping concentration
$0<\rho<\rho_c$, since all these doping concentrations correspond to the same value
of the chemical potential $\mu=\mu_c$ at which the ground state energy (\ref{energy})
is $SO(5)$ symmetric. 

However, it should be pointed out that the projected $SO(5)$ symmetry at 
$\mu=\mu_c$ has only been demonstrated within the variational mean field
approximation. This corresponds to the semiclassical limit, and becomes
exact only in the large $s$ limit, where $s$ labels the representation
of the local $SO(5)$ group at a given site. Quantum fluctuations 
can be systematically investigated
as a $1/s$ expansion. Assuming uniform ground states, we have studied the
effect of zero point fluctuations in Sec. \ref{fluctuations}
and found that at $\mu=\mu_c$, 
the intermediate mixed states have slightly higher energy than the AF and 
SC state. Therefore, quantum fluctuation leads to a slight breaking of the
projected $SO(5)$ symmetry. The important point here is that this symmetry
breaking effect can be systematically controlled in the semi-classical $1/s$
expansion, and certainly is independent of the Coulomb energy scale $U$.
This fluctuation would induce a first order transition and 
predict phase separation of AF and SC states at
$\mu=\mu_c$. However, there are also other competing interactions such
as nearest-neighbor and next-nearest-neighbor interactions which tend to
reduce the barrier, and could also lead to non-uniform states like stripes.
Due to the complexity of the calculations, we
shall defer the detailed studies of these competing effects to future works. 

\section{Phase Diagram}
\label{phase}

\begin{figure*}[h]
\centerline{\epsfysize=10.0cm 
\epsfbox{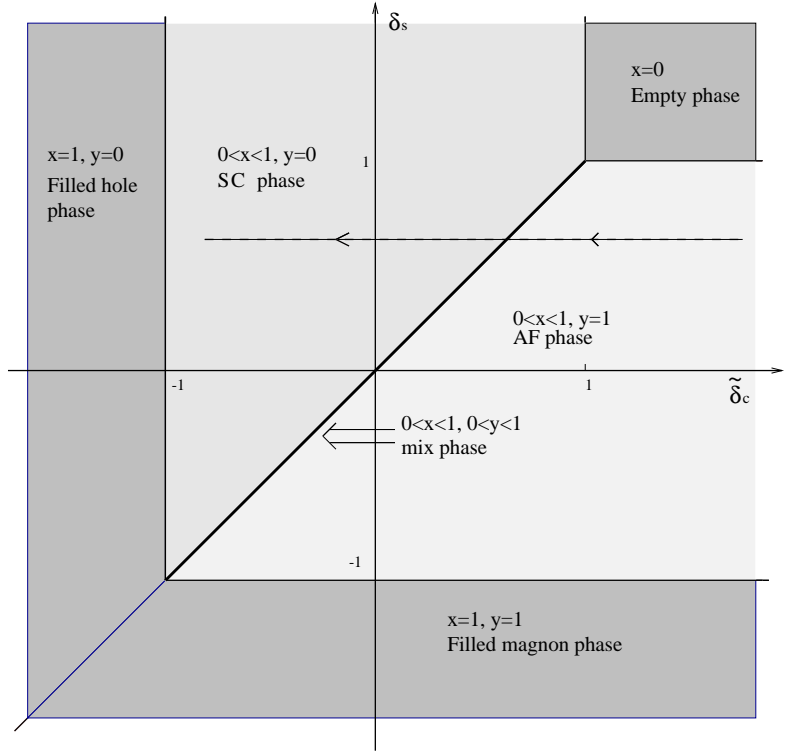}
}
\caption{Phase diagram of the projected $SO(5)$ model in the $\delta_s$ versus
$\tilde\delta_c$ plane. Phase boundaries are depicted by the solid lines.
Variation of the chemical potential traces out a 
one dimensional trajectory as shown on the dotted line.}
\label{fig2}
\end{figure*}

In this section we investigate the phase diagram of the projected $SO(5)$ model
within the framework of the variational wave function (\ref{variational}). 
Taking uniform values of the variational parameters $\theta$, 
$m_x=\cos\alpha$, $m_1=\sin\alpha$ and $m_y=m_z=m_5=0$, the variational energy
in (\ref{energy}) reduces to
\begin{equation}
\frac{E(\theta,\alpha)}{N} = -J_s \sin^2 2\theta \cos^2\alpha
-\frac{J_c}{2} \sin^2 2\theta \sin^2\alpha + \Delta_s \sin^2 \theta \cos^2\alpha
+\tilde\Delta_c \sin^2 \theta \sin^2\alpha
\end{equation}
In the following, we shall mainly study the $SO(5)$ symmetric case, and
take $J_c=2J_s=2J$. Defining $x=\sin^2\theta$ and $y=\cos^2\alpha$, and the
dimensionless coupling constants $\epsilon\equiv \frac{E(x,y)}{4JN}$,
$\delta_s\equiv \frac{\Delta_s}{4J}$ and 
$\tilde\delta_c\equiv \frac{\Delta_c}{4J}$, we obtain
\begin{equation}
\epsilon(x,y) = x^2-x+\tilde\delta_c x+
(\delta_s-\tilde\delta_c)xy
\label{mean_energy}
\end{equation}
We shall minimize (\ref{mean_energy}) with respect to $x$ and $y$, subject to
the condition that $0\leq x,y\leq 1$. 
The phase diagram can be plotted in the two dimensional parameter
space of $\tilde\delta_c$ and $\delta_s$. We 
notice that $\epsilon(x,y)$ depends linearly on $y$, therefore,
for $\delta_s > \tilde\delta_c$ we obtain $y_{min}=0$ and 
\begin{eqnarray}
& 0<x_{min}=\frac{1-\tilde\delta_c}{2}<1 
& \ \ \ for \ \ \  -1<\tilde\delta_c < 1 \nonumber \\
& x_{min}=0 & \ \ \ for \ \ \ \tilde\delta_c > 1    \nonumber \\
& x_{min}=1 & \ \ \ for \ \ \ \tilde\delta_c < -1
\end{eqnarray}
Similarly, for $\delta_s < \tilde\delta_c$ we obtain $y_{min}=1$ and 
\begin{eqnarray}
& 0<x_{min}=\frac{1-\delta_s}{2}<1 
& \ \ \ for \ \ \ -1< \delta_s < 1 \nonumber \\
& x_{min}=0  & \ \ \ for \ \ \  \delta_s > 1    \nonumber \\
& x_{min}=1  & \ \ \ for \ \ \ \delta_s < -1
\end{eqnarray}
 From these equation, we can determine the phase diagram as shown in Fig. (\ref{fig2}).
There are seven different phases on this phase diagram.  
$x_{min}=0$ corresponds to a quantum disordered
singlet state with no condensed bosons.  $x_{min}=1$ and $y_{min}=0$ corresponds
to a quantum disordered state with completely filled hole pairs. 
$x_{min}=1$ and $y_{min}=1$ corresponds
to a quantum disordered state with completely filled magnons.
$0<x_{min}<1$ and $y_{min}=1$ describe a pure AF phase, while 
$0<x_{min}<1$ and $y_{min}=0$ describes a pure SC phase. 
When $-1< \delta_s =\tilde\delta_c < 1$ 
a continuous family of mixed AF/SC states labeled by a free superspin angle
$0<\alpha<\pi/2$ is realized, while for
$\delta_s =\tilde\delta_c < -1$ 
a continuous family of quantum disordered states labeled by a free superspin angle
$0<\alpha<\pi/2$ is obtained.

\begin{figure*}[h]
\centerline{\epsfysize=4.0cm 
\epsfbox{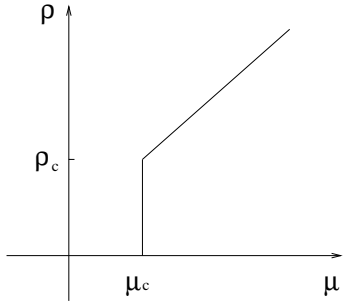}
}
\caption{Density versus chemical potential relation in the projected 
$SO(5)$ model. Unlike the case of a generic first order transition,
the ground state in the density range $0<\rho<\rho_c$ is uniform,
rather than phase separated.}
\label{fig3}
\end{figure*}

The system traces out a one dimensional trajectory in this two dimensional
phase diagram as the chemical potential is increased, as depicted in
Fig. (\ref{fig2}). 
Increasing the chemical potential decreases the $\tilde\delta_c$ parameter
while holding $\delta_s$ constant. $\delta_s$ describes the degree
of quantum spin fluctuations in the system, since in the AF phase, the
size of the Neel moment
\begin{equation}
m_{AF}=\sqrt{\frac{1}{2}(1-\delta_s^2)}
\label{moment}
\end{equation}
decreases with increasing $\delta_s$.
For $\delta_s<1$, the system goes through
a phase transition from AF to SC at $\mu=\mu_c=\Delta_c-\Delta_s$. At this 
critical value of the chemical potential, the $\theta$ parameter remains 
fixed, but the $\alpha$ parameter changes continuously from $0$ to $\pi/2$,
and correspondingly, the density changes from $0$ to 
\begin{equation}
\rho_c\equiv \sin^2\theta=x_{min}=\frac{1-\delta_s}{2}
\label{rho_c}
\end{equation}
This behavior gives a density versus chemical potential diagram as shown in
Fig. (\ref{fig3}). 
As we see, for densities in the range $0<\rho<\rho_c$, the 
system is infinitely compressible since $\partial\rho/\partial\mu=\infty$.
For $\mu>\mu_c$ or $\rho>\rho_c$, the system has a finite compressibility
of $\partial\rho/\partial\mu=1/8J$.

\begin{figure*}[h]
\centerline{\epsfysize=6.0cm 
\epsfbox{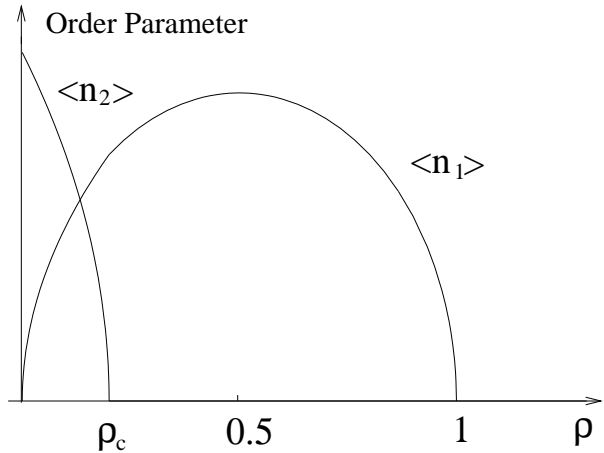}
}
\caption{AF ($\langle n_2\rangle$) and SC ($\langle n_1\rangle$)
order parameters versus density in the 
projected $SO(5)$ model.}
\label{fig4}
\end{figure*}

It is interesting to plot both the AF and the SC order parameters 
$\langle n_1\rangle$ and $\langle n_2\rangle$ as a function of the density
for the whole range of $0<\rho<1$. We will restrict to the case of 
$0<\delta_s<1$ where the undoped state is a AF state. We obtain the following
doping dependence of the SC order parameter:
\begin{equation}
\langle n_1\rangle = \left\{ \begin{array} {l}
    \sqrt{\rho(1-\rho_c)} \ \ \ for \ \ \ \rho<\rho_c \\
    \sqrt{\rho(1-\rho)}   \ \ \ for \ \ \ \rho>\rho_c
  \end{array} \right. 
\end{equation}
and the doping dependence of the of the AF order parameter:
\begin{equation}
\langle n_2\rangle = \left\{ \begin{array} {l}
    \sqrt{2(1-\rho_c)(\rho_c-\rho)} \ \ \ for \ \ \ \rho<\rho_c \\
    0   \ \ \ for \ \ \ \rho>\rho_c
  \end{array} \right. 
\end{equation}
These behaviors are depicted in Fig. (\ref{fig4}).

We note several interesting features of the phase diagram. First of all,
we can use the energy functional (\ref{energy}) as a starting point for
a finite temperature classical fluctuation analysis, and estimate the
transition temperature due to the classical fluctuations. 
Within such a framework, the 3D AF ($T_N$) and SC ($T_c$) transition
temperatures are proportional to the stiffness of the spin and the 
phase fluctuations, which are in turn proportional 
to $\langle n_2\rangle^2$ and $\langle n_1\rangle^2$ 
respectively. Therefore, Fig. (\ref{fig4}) gives an approximate estimate of the
transition temperatures. For small $\rho_c$, we see that there is
a sharp drop of $T_N$ and a maximum of $T_c$ at $\rho=1/2$. 
The system is a AF insulator at doping $\rho=0$ and a pure SC 
state for $\rho>\rho_c$. The reason for a maximum of $T_c$ at 
$\rho=1/2$ is due to the strong correlation of the charged bosons. 
The charge bosons have a hard-core
interaction, therefore, they are insulating at both $\rho=0$ and 
$\rho=1$ and have maximal charge stiffness at $\rho=1/2$. We can
perform a rough translation of this optimal doping value in our
effective model to the microscopic model. Since our effective
model is defined on a unit cell with twice the lattice spacing
of the microscopic model, $\rho=1/2$ therefore describes one
hole pair per four sites in the microscopic model, or a doping
of $x=1/4=25\%$ in the conventional language. This crude 
argument tends to overestimate the value for optimal doping,
since it neglects the effects of unpaired electrons. However,
considering the crudeness of the estimate, it is still
reasonably close to the optimal doping $x=15\%$ observed in the LSCO
family of high $T_c$ superconductors.

In the regime of $0<\rho<\rho_c$, the system is a coherent mixture 
of AF and SC order. For this
entire range of densities, the system has a projected $SO(5)$
symmetry within the variational approximation discussed above. 
The projected $SO(5)$ symmetry manifests itself in terms of 
a infinite compressibility in the region $0<\rho<\rho_c$
and, as we shall see in next section, 
a charge mode with a dispersion relation
$\omega\sim k^2$. Since such a state is rather
unusual, and maybe highly susceptible to density fluctuations,
we would like to discuss more detailed physical properties in
this region.

First let us comment on the fact that there are familiar physical
systems whose uniform ground states are infinitely compressible.
The free boson model is certainly such an example, and the density
mode also has a $\omega\sim k^2$ dispersion relation. But the
infinite compressibility is due to the absence of the interaction,
which is not characteristic of the strongly interaction system
considered here. A less trivial example is the spin $1/2$ XXZ ferromagnetic
Heisenberg model, given by the Hamiltonian,
\begin{equation}
{\cal H} = J \sum_{i,j} (S^x_i S^x_j + S^y_i S^y_j + \Delta S^z_i S^z_j)
\end{equation}
where $J<0$ and the sum extends over nearest neighbor sites of a square lattice. 
This model can be interpreted as quantum hard-core boson model, where
the fully polarized spin down state could be identified with the
vacuum of the bosons, the XY part of the Hamiltonian describes the
hopping of the bosons and the last term describes the nearest neighbor
attraction between the bosons if $\Delta>0$. When $\Delta>1$, the system
is in the Ising limit, and the spontaneous breaking of the $Z_2$ symmetry
implies phase separation of the bosons. On the other hand, when 
$0<\Delta<1$, the system is in the XY limit, and the ground state is 
a superfluid. Therefore, the anisotropy parameter describe the
competition between superfluidity and phase separation. At $\Delta=1$,
the system has a $SU(2)$ symmetry and the dispersion relation becomes
quadratic. Different directions of the
ferromagnetic polarizations are degenerate and
can be changed without any energy cost.
Since the $z$ component of the ferromagnetic polarization is identified
with the total density of the hard-core bosons, the $SU(2)$ vacuum
degeneracy implies infinite compressibility of the corresponding boson
system for the entire range of boson densities $0<\rho<1$. 

These two examples illustrates that there is nothing intrinsically
pathological about having a system with infinite compressibility.
The second example is more generic, and shows that uniform states
with infinite compressibility can be obtained in systems on the
verge of phase separation, and the infinite compressibility can be
ensured by symmetry. Both of these properties are also shared by
the projected $SO(5)$ model. These models are on the verge of 
phase separation into AF and SC phases, and the infinite compressibility
is a result of the projected $SO(5)$ symmetry.
It indicates that
small perturbations, such as next nearest neighbor interactions, quantum fluctuations,
and quenched disorder will be very important to determine the true ground state.
With such perturbations, the ground state is expected to be unstable toward
the experimentally reported textures
such as the spin glass, stripes and incommensurate spin density waves.

Next let us investigate the phenomenological consequence of this
remarkable property. One of the most puzzling 
properties of the high $T_c$ superconductors is the constant
chemical potential in the underdoped samples. For LSCO systems,
where the doping level can be varied continuously by the $Sr$
concentration, this effect has been dramatically observed in
the ARPES experiments and the constant chemical potential 
persists from the weakly doped insulator to optimally doped
superconductor\cite{fujimori}. In fact numerical calculations on the
Hubbard model also reveal similar divergent behavior of the 
compressibility as the metal-insulator transition is approached from
the metallic side\cite{imada}.

The simplest explanation of the small chemical potential shift
is a two phase mixture with different densities at a first order 
phase transition. If the system globally phase separates into
two different spatial regions with different charge densities 
but the same free energy densities, the added charges only change
the proportion of mixture of the two phases and do not change
the energy, therefore, $\partial\mu/\partial\rho=0$. However,
this situation of global phase separation can certainly not
occur in a system with long ranged Coulomb interaction and is
ruled out in the real high $T_c$ system.

A phenomenon possibly related to the tendency of phase separation
is the formation of stripes\cite{zaanen,kivelson,scalapino2}. 
A stripe state can be viewed as 
microscopic phase separation of AF and SC into alternating regions,
where each region has different charge density and the same free energy
density. However, a crucial difference between the global phase
separation and this picture of microscopic phase separation is that
the stripe state has infinitely many surfaces between AF and SC,
and the surface energy makes a finite contribution to the total
energy in the thermodynamic limit. In this picture, doping can be
accomplished by converting AF stripes into SC stripes, thereby 
creating more surfaces separating AF and SC regions. Finite 
doping density therefore leads to a finite density of surfaces
and the accumulated surface energy would in general lead to 
a shift of the chemical potential {\it under a generic situation}.
Additional physical conditions are needed to ensure the constant
chemical potential in the stripe phase.

Therefore, the absence of the chemical potential shift 
places a very strong constraint on possible theoretical
explanations. The projected $SO(5)$ model proposed in this work
offers a possible explanation for the absence of chemical
potential shift. At a critical value of the chemical potential
where the AF and SC states have degenerate energy density,
we can have three situations, where
the intermediate mixed states have higher, lower or degenerate
energy compared to the AF and SC states. When the intermediate
states have higher energy, the system will go through a first
order phase transition at $\mu=\mu_c$
and this will lead to global phase separation into AF and SC regions. 
On the other hand, if the intermediate states have lower energy, there
exists a range of chemical potential $\mu_{c1}<\mu<\mu_{c2}$ where
the mixed phase has a uniform and continuously varying density. 
In this case, $\partial\mu/\partial\rho\ne 0$ is obtained.
When the region $\mu_{c1}<\mu<\mu_{c2}$ shrinks to zero, we obtain  
the limiting $SO(5)$ symmetric case where the system is on the boundary
between a first order transition and two second order phase transitions.
In this case, $\partial\mu/\partial\rho=0$ for a range
of densities $0<\rho<\rho_c$. Therefore, if we restrict to 
ground states where the density is not globally inhomogeneous,
the absence of the chemical shift directly implies
$SO(5)$ symmetry.

Within this model, we can therefore define a experimental procedure
to measure one of the most crucial parameter of the theory,
namely $\rho_c$. Using the experimental ARPES data for LSCO system
we would identify $\rho_c$ to be approximately the same as the 
optimal doping density. We recall that $\rho_c$ is also a measure
of the degree of the quantum spin fluctuation in the system. 
For the bi-layer materials such as YBCO and BISCO superconductors,
the quantum spin fluctuation are stronger due to the inter-layer
spin exchange, and we would predict that $\rho_c$ should be less
than the optimal doping value.

\section{Collective modes}
\label{modes}

Having discussed the ground state properties and the phase diagram of the model,
we are now in a position to study the collective excitations of the model. We have
argued that the ground state energy can remain $SO(5)$ symmetric despite the 
projection. However, the projection does affect the collective
excitation spectrum near half-filling. Nonetheless, as we shall see, there remains
a unique signature of the projected $SO(5)$ symmetry in the collective excitation
spectra.

In principle, the collective excitation spectra can be obtained straightforwardly
by studying the quadratic fluctuations around the mean field minima. The resulting
quadratic boson Hamiltonian can be simply diagonalized. The main complication in the
procedure is the hard-core boson constraint, which requires
\begin{equation}
t^\dagger_\alpha t_\alpha (x) + t^\dagger_h t_h (x) \leq 1
\label{hard-core}
\end{equation}
for every site. There are several ways to implement this constraint rigorously.
One is to follow the mapping from the one-component hard-core
boson model to the $XY$ model and generalize it to a multi-component hard-core boson
model. One could also convert the above inequality constraint to an equality
constraint by introducing a boson creation and annihilation operator for the
singlet state. This approach will be implemented in the Appendix.
For simplicity of presentation, here we shall adopt a less rigorous approach and
introduce a on-site boson repulsion term 
\begin{equation}
W \sum_x (t^\dagger_\alpha t_\alpha+ t^\dagger_h t_h)^2
\label{soft-core}
\end{equation}
to our Hamiltonian (\ref{Hamiltonian}) and
convert the hard-core constraint to a soft-core
constraint. We shall show later that all results can be expressed in terms
of the order parameter, which is implicitly dependent on $W$, but there is
no explicit dependence on $W$.
We have verified that all three methods give the same long wave length
spectra for the collective modes in the limit of low boson density. 
In the next section, we shall present another
calculation based on the continuum effective Lagrangian method, which also reproduces
the same spectra. 

To simplify presentation, we shall concentrate on the case where the ground
state energy functional (\ref{energy}) is $SO(5)$ symmetric, {\it i.e.} for coupling
constants $J_c=2J_s\equiv 2J$ and $\tilde\Delta_c=\Delta_s\equiv\Delta$. We choose
the direction of spontaneously broken symmetry to be 
$\langle t_x\rangle = \langle t^\dagger_x\rangle = x$ and
$\langle t_h\rangle = \langle t^\dagger_h\rangle = y$. The extremal condition
can be easily determined to be
\begin{equation}
x^2+y^2 \equiv r^2 = \frac{4J-\Delta}{2W}
\label{extremal}
\end{equation}
The combination $x^2+y^2$ expresses the fact that the classical minimum is 
$SO(5)$ symmetric. We can therefore write $x=r \cos\alpha$ and $y=r \sin\alpha$.
Expanding the boson operators as:
\begin{equation}
t_x=x+a_x, \ \ t_y=a_y, \ \ t_z=a_z, \ \ t_h=y+a_h   
\label{expansion}
\end{equation}
we obtain the following quadratic Hamiltonian
\begin{eqnarray}
 H &=& (\Delta +2Wr^2)\sum_x(a_{x}^{\dagger}a_x+a_{h}^{\dagger}a_h) 
   + Wx^2\sum_x (a_{x}^{\dagger}+a_x)^2+ Wy^2\sum_x (a_{h}^{\dagger}+a_h)^2
   \nonumber \\
   &-& \frac{J}{2}\sum_{<x,x'>} (a_{x}^{\dagger}(x)+a_x(x))( a_{x}^{\dagger}(x')+a_x(x')) 
  -J\sum_{<x,x'>}(a_{h}^{\dagger}(x)a_h(x')+h.c.)
  \nonumber \\  
  &+& 2Wxy \sum_{x}( a_{h}^{\dagger}(x)+a_h(x))( a_{x}^{\dagger}(x)+a_x(x))
  \nonumber \\
  &-& \frac{J}{2}\sum_{<x,x'>}( a_{y}^{\dagger}(x)+a_y(x))( a_{y}^{\dagger}(x')+a_y(x')) 
  - \frac{J}{2}\sum_{<x,x'>}( a_{z}^{\dagger}(x)+a_z(x))( a_{z}^{\dagger}(x')+a_z(x')) 
  \nonumber \\
  &+& (\Delta +2Wr^2)\sum_x(a_{y}^{\dagger}a_y+a_{z}^{\dagger}a_z)
\label{quadratic}
\end{eqnarray}
We are in particular interested in the collective mode spectra for the 
AF insulating state with $\alpha=0$, the mixed states with $0<\alpha<\pi/2$
and the SC state with $\alpha=\pi/2$ and how they connect to each other.

 From this quadratic Hamiltonian we can learn a number of important features.
First we notice that the $a_y$ and $a_z$ modes 
are decoupled for all ranges of $0\leq \alpha\leq \pi/2$, but most 
importantly, their dispersion relations are independent of $\alpha$
and given by 
\begin{equation}
\omega(k) = v_s k \ \ , \ \ v_s = 2J
\label{spin_wave}
\end{equation}
where $k\equiv a |\vec k|$ and $a$ is the lattice constant.
This is indeed a very remarkable property. At $\alpha=0$, $a_y$ and 
$a_z$ modes are nothing but the transverse AF spin
wave modes. AF spin waves are usually viewed
as Goldstone modes and their existence is due to the AF long range order.
However, as $\alpha$ changes continuously from $0$ to $\pi/2$, the AF long
range order continuously diminishes until it vanishes at $\alpha=\pi/2$.
The reason that the properties of the $a_y$ and $a_z$ modes do not change
at all is due to the $SO(5)$ symmetry of this model, since the
diminishing AF order is compensated by the increasing SC order as the superspin
angle $\alpha$ is varied. As we shall see, the $a_x$ mode is the AF spin
amplitude mode at $\alpha=0$, but it becomes massless and degenerate with
the $a_y$ and $a_z$ modes at $\alpha=\pi/2$. These three modes form a massless
$\pi$ triplet mode whose existence is purely a consequence of the SC order.
Therefore, as $\alpha$ is continuously varied from $0$ to $\pi/2$, the
transverse AF spin wave modes gradually change their character to become
the $\pi$ triplet resonance of the SC state. As we shall see, for $\mu>\mu_c$,
the $\pi$ triplet mode becomes massive.

At the AF point $\alpha=0$, the spin amplitude mode $a_x$ is decoupled from the
SC mode $a_h$ and can be diagonalized separately. The dispersion for the spin 
amplitude mode has the conventional massive relativistic form    
for small $k$ 
\begin{equation}
\omega^2(k) = 16J^2 (k^2/4 + m_x^2) \ \ , \ \ m_{x}^{2} = \frac{Wx^2}{J} = 
\frac{4J-\Delta}{2J}
\label{spin_amp}
\end{equation}
On the other hand, we have a massless SC Goldstone mode $a_h$ with the
following dispersion, 
\begin{equation}
\omega(k) = J k^2
\label{SC-mode}
\end{equation}
This mode is an important new prediction of the $SO(5)$ theory. It is the 
counterpart of the $\pi$ resonance in the AF state. In the unprojected
$SO(5)$ model, there are two such modes, with charge $\pm 2$, and they 
represent gapless fluctuations from AF to SC at half-filling. In the
projected $SO(5)$ model, the charge $+2$ mode is projected out of the spectrum,
however, the charge $-2$ mode remain massless at $\mu=\mu_c$.  It is 
also a manifestation of the gapless fluctuation from AF to SC at half-filling, 
but the SC fluctuation is hole-like, rather than both hole and particle-like 
as in the unprojected case. We see again that a large Mott-Hubbard gap is
fully compatible with gapless SC fluctuation at half-filling.
Experimental detection of this mode could provide a important test of the
projected $SO(5)$ symmetry.
 
The $a_x$ and the $a_h$ modes also decouple in the pure SC state with 
$\alpha=\pi/2$. However, their physical interpretation change.
The $a_x$ mode becomes gapless at this point with the same dispersion
as in (\ref{spin_wave}). Therefore, the three modes $a_x$, $a_y$ and $a_z$
form a gapless $\pi$ triplet mode of the pure SC state, and represent
the gapless fluctuation from SC to AF at $\mu=\mu_c$, but with a finite
hole density $\rho_c$ given in (\ref{rho_c}). 
The dispersion for the $a_h$ mode is given by
\begin{equation}
\omega(k) = 2J m_h k \ \ , \ \ m_{h}^{2} = \frac{Wy^2}{J} =
\frac{4J-\Delta}{2J} 
\label{phase_mode}
\end{equation}
and has the natural interpretation of a linearly dispersing phase Goldstone
mode of the SC state.

In the intermediate mixed state with $0<\alpha<\pi/2$, the $a_x$ and the
the $a_h$ modes are coupled. Diagonalization of these modes gives:
\begin{equation}
\frac{\omega^2(k)}{(4J)^2} = \left\{ \begin{array}{l}
    (1+m_h^2)k^2/4+m_x^2 \\
    (1+m_h^2/m_x^2)k^4/16 
 \end{array} \right. \ \ \
\left\{ \begin{array}{l}
 m_x^2=\frac{4J-\Delta}{2J}\cos^2\alpha \\
  m_h^2=\frac{4J-\Delta}{2J}\sin^2\alpha 
 \end{array} \right. 
\label{mixed_modes}
\end{equation}
The upper massive mode has predominantly spin amplitude character,
and we see that the gap diminishes continuously until it reaches
zero at $\alpha=\pi/2$ to become the massless $\pi$ triplet mode.
The lower mode has predominantly SC fluctuation character, and has
a gapless $\omega\propto k^2$ dispersion. In the mixed region where
both the AF and the SC order parameters are non-zero, one would 
naturally expect a gapless phase mode corresponding to the SC order.
However, in a interacting boson system, the phase mode is expected
to have linear dispersion on general ground. Therefore, what is 
surpassing and new here is not the gapless nature of the SC mode,
but its {\it quadratic dispersion}. In order to locate the origin of the
quadratic dispersion, we have perturbed the model away from the
projected $SO(5)$ symmetric point so that a uniform 
mixed state is stabilized as
a classical minimum. SC fluctuation around such a non-$SO(5)$ symmetric
point is gapless and has linear dispersion. A quadratic dispersion
is only realized at the $SO(5)$ symmetric point. Therefore, the
quadratic dispersion is a unique signature of the projected $SO(5)$
symmetry in the entire range of densities $0<\rho<\rho_c$! 
To understand the physical origin of this
remarkable phenomenon, we notice that a boson system with gapless 
quadratic dispersion generally has infinite compressibility. This 
can be directly seen from the compressibility sum rule
\begin{equation}
\kappa \equiv \frac{\partial\rho}{\partial\mu} 
= \frac{1}{2} \lim_{k\rightarrow 0}\lim_{\omega\rightarrow 0} 
  \chi(k,\omega) \sim 
  \lim_{k\rightarrow 0} \frac{k^2}{\omega^2(k)}
\label{compressibility}
\end{equation}
where $\chi(k,\omega)$ is the dynamical density correlation function.
Because of the quadratic dispersion relation, we can see explicitly that
$\chi(k)\propto 1/k^2$ for small $k$, therefore, a infinite compressibility
is obtained. On the other hand, a infinite compressibility implies that
$\frac{\partial\mu}{\partial\rho}=0$, {\it i.e.} the chemical potential
is independent of doping. But this is exactly the prediction of the 
projected $SO(5)$ model! For $0<\rho<\rho_c$, the chemical potential is 
pinned at the $SO(5)$ symmetric point $\mu=\mu_c$, where the
superspin vector can point in any direction. To accommodate a
long wave length fluctuation of the hole density, the system rotates 
into another degenerate minimum with a different superspin angle $\alpha$
and a different hole density. 
For this reason, the chemical potential does not change and the system
is infinitely compressible.

\section{Quantum corrections to the mean-field solution}
\label{fluctuations}

At the mean-field level,
the ground-state energy of the Hamiltonian
(\ref{Hamiltonian}) with (\ref{soft-core})
depends on the AF and SC order parameters $x$ and $y$ only via
their combination $r^2=x^2+y^2$ reflecting the SO(5) invariance of the
mean-field result.
However, the zero-point energy of the bosons in the
quadratic Hamiltonian gives a correction to the
ground-state energy due to quantum fluctuations. (Calculations of the quantum
fluctuation effects at the $SO(3)$ spin-flop transition have been
studied in ref. \cite{assa2}.
As we will show below, this correction turns out to depend on $x^2$
and $y^2$ separately. 

For the soft-constraint case, $E_1$ can be expressed as
\begin{equation}
\label{egs}
E_1 = \frac12 \sum_k
   \sum_{i} \omega_i(k) - 2 (\Delta + 2 W r^2)
\end{equation}
where $\omega_i(k)$ are the four collective modes described in
Sec. \ref{modes} (although extended to all values of $k$ in the BZ).
 In a systematic $1/s$ expansion, this correction
to the ground state energy scales like $1/s$, and is therefore small
in the semi-classical limit. Alternatively,
$E_1$ can be seen as a small correction to the mean-field energy $E_0$
for small values of the parameter
$\epsilon \equiv  \frac{r^2 W}{2 J}=1-\frac{\Delta}{4J}$.
However, contrary to $E_0$, $E_1$ also depends on the superspin angle
$\alpha$ and thus produces a small  SO(5)-symmetry breaking.

We now evaluate the $\alpha$-dependent part of Eq. (\ref{egs}) in the
small-$\epsilon$ limit.
We thus parametrize
$\chi = 1- 2 \sin^2\alpha$ ($-1 \leq \chi \leq 1$), 
 differentiate
 the expression (\ref{egs}) with respect to $\chi$, and expand it 
 up to second order in $\epsilon$.
By further transforming $(\cos k_x + \cos k_y)/2=1-q$ with $ 0\leq q\leq 2$,
the derivative of $E_1$ can be expressed as the sum of two terms
\begin{equation}
\frac{d}{d \chi} E_1 = E_{1A}'  + E_{1B}' \;,
\end{equation}
where
\begin{equation}
\label{intg}
{E_{1A}'\atop E_{1B}'}  = 2 J \int_0^2 d\  q \ {\cal D}( q) 
{G_{A} \atop G_{B}} \;,
\end{equation}
with
\begin{equation}
\label{g1}
G_A \equiv 
=
  {\frac{\left( \left(  {1-\sqrt{ q}} \right) \,
         \epsilon \right) }{2\,{\sqrt{ q}}}} +O(\epsilon^2)\;,
\end{equation}
\begin{equation}
\label{g2}
G_B \equiv 
- 
   {\frac{\left( {\sqrt{ q}-1} \right) \,
       \left(  \chi\,{\sqrt{ q}}  -1 - \chi - {\sqrt{ q}}  
         \right) \,
       \epsilon^2}{4\,
       \left( 1 + {\sqrt{ q}} \right) \,
       {{ q}^{{\frac{3}{2}}}}}} + O(\epsilon^3) \;,
\end{equation}
and the density of states ${\cal D}(q)$ is defined as
\begin{equation}
{\cal D}(q) = \frac{1}{4 \pi^2}
\int \ d k_x \ d k_y \ \delta\left[q-1+(\cos k_x + \cos
  k_y)/2\right] \;.
\end{equation}
Unfortunately, 
 the terms of the $\epsilon$ expansion in $G_B$ diverge 
 when integrated over $ q$. 
It is thus convenient to carry out the transformation $q= \epsilon z$ in
$G_B$ and {\it then} expand in powers of $\epsilon$.
We obtain
\begin{equation}
\label{g2u}
G_B = \frac{\sqrt{\epsilon}}{2} \left( \frac{1}{\sqrt{z + 1 + \chi }}
- \frac{1}{\sqrt{z}} \right) + O(\epsilon) \;.
\end{equation}
For small $\epsilon$,
the integral in $z$ can be extended
to $\infty$ 
and we obtain from Eq. (\ref{intg})
\begin{equation}
\label{e1bf}
E_{1B}' = 2J \ 
\epsilon \int_0^{\infty} d z\   {\cal D}(z \epsilon)\ G_B = 
\frac{ -(\sqrt{1 + \chi }\ \epsilon^{3/2})}{\pi} + 
O(\epsilon^2 \log \epsilon)\;.
\end{equation}
While in evaluating Eq. (\ref{e1bf}) we only need
 the density of states at $q=0$,
${\cal D}(0) = \pi^{-1}$, 
for the  first term $E_{1A}'$ one needs ${\cal D}(q)$
in the whole domain $0 \leq q \leq 2$ 
However,  $E_{1A}'$ is independent of $\chi$ and thus it merely fixes
the value of the critical chemical potential.
A numerical integration yields 
\begin{equation}
\label{e1af}
E_{1A}' = 0.28 \ J \ \epsilon \;.
\end{equation}
By integrating over $\chi$ Eqs. (\ref{e1af}) and (\ref{e1bf}), we finally obtain the total contribution
to the ground-state energy correction.
\begin{equation}
\frac{E_1}{2 J} = 0.14\ \chi \ \epsilon   
-\frac{2}{3 \pi} \left(\epsilon\ (1 + \chi)\right)^{3/2} + O(\epsilon^2 \log \epsilon) +
{\rm const.}\;.
\end{equation}
$E_1$ thus lifts the degeneracy as a function of $\chi$ and initially
   favors
the pure superconducting phase ($\chi=-1$).
A small chemical potential term 
$-\mu' y^2 = -\mu' r^2 (1-\chi)/2 $ with
$\frac{\mu' r^2}{2 J}= \frac{\mu'_c r^2}{2 J}= - 0.28 \ \epsilon + 
\frac{2}{3 \pi} \ (2 \epsilon)^{3/2} $
restores the degeneracy between the
pure superconducting  ($\chi=-1$) and the pure antiferromagnetic
($\chi=+1$) phases. However, due to the convexity of $E_1$ as a function of
   $\chi$ there is always a barrier ($\propto J \epsilon^{3/2}$)
between the two phases since
the mixed phase always has a higher energy. This means that
at $\mu'=\mu'_c$ and for intermediate densities
the system prefers to phase separate
between the two pure phases
rather than choosing the mixed phase.
However, the important point is that this barrier remains small in the
limit $U\to\infty$.

In view of the symmetry breaking effects of the quantum fluctuations,
it would be interesting to see whether there is a limit when the
the wave function (\ref{variational}) and the projected $SO(5)$ symmetry 
becomes exact. Rokhsar and Kotliar\cite{rokhsar}
have shown that this type of wave functions
are actually exact in the limit of infinite dimensions. Therefore,
besides the $1/s$ expansion, we could also use a $1/d$ expansion
(where $d$ is the space dimension) to
systematically control the $SO(5)$ symmetry breaking quantum effects.
Besides quantum fluctuations, there are also other symmetry breaking terms.
Nearest-neighbor and next-nearest-neighbor Coulomb interactions also break
the $SO(5)$ symmetry, however, their corrections to the ground state is
concave, {\it i.e.} the energy of the intermediate states are lowered.
 Therefore,
 they can also lead to uniform mixed states in some region of the
phase diagram. The detailed study of all these competing effects will
be carried out in subsequent works.

\section{Low energy effective Lagrangian}
\label{lagrangian}
While the projected $SO(5)$ model defined on a lattice enables us to make
some connection to the underlying microscopic physics, for most discussions
concerning the long wave length and low energy degrees of freedom, it is
desirable to have a effective continuum Lagrangian. Such a formulation can be directly
obtained by taking the long wave length limit of the projected bosonic model
discussed previously. However, in order to make the connection to the 
unprojected $SO(5)$ model clearer, we shall motivate our discussion from
the original $SO(5)$ effective model.

The effective Lagrangian for a fully $SO(5)$ symmetric model takes the form of
\begin{equation}
{\cal L} = \frac{\chi}{2} (\partial_t n_a)^2 - \frac{\rho}{2} (\partial_k n_a)^2
- V(n) 
\label{full_L}
\end{equation}
where $\chi$ measures the superspin susceptibility, $\rho$ measures the
superspin stiffness, $k=x,y$ denotes the spatial directions and $V(n)$ 
is a scalar function of the superspin magnitude $n_a^2$ only. There are
three important symmetry breaking effects connected with the presence of 
a large Mott-Hubbard gap. First is an asymmetry in the scalar potential,
which can be described by a additional term 
\begin{equation}
V_g(n) = + \frac{g}{2} (n_1^2 + n_5^2)
\label{V_g}
\end{equation}
which for positive $g$ favors AF at half-filling. Second is the asymmetry 
between the spin ($\chi_s$) and the charge ($\chi_c$) susceptibilities, which
modifies the kinetic energy to 
\begin{equation}
\frac{\chi_s}{2} (\partial_t n_\alpha)^2 + \frac{\chi_c}{2} (\partial_t n_i)^2 
\end{equation}
The last symmetry breaking effect is due to the chemical potential $\mu$, which
enters the Lagrangian as a gauge coupling in the time direction, and modifies the
charge part of the kinetic energy to 
\begin{equation}
\frac{\chi_c}{2} ((\partial_t n_1 + \mu n_5)^2 + (\partial_t n_5 - \mu n_1)^2)
\end{equation}
Combining these three symmetry breaking terms, we obtain
\begin{eqnarray}
{\cal L} &=& \frac{\chi_s}{2} (\partial_t n_\alpha)^2 +
\frac{\chi_c}{2} ((\partial_t n_1 + \mu n_5)^2 + (\partial_t n_5 - \mu n_1)^2) -
\frac{\rho}{2} (\partial_k n_a)^2
- V(n) - \frac{g}{2}n_i^2   \nonumber \\
&=& \frac{\chi_s}{2} (\partial_t n_\alpha)^2 + \frac{\chi_c}{2} (\partial_t n_i)^2
+ \mu\chi_c(n_5\partial_t n_1 - n_1\partial_t n_5) + \frac{\mu^2\chi_c-g}{2} n_i^2 
-\frac{\rho}{2} (\partial_k n_a)^2 - V(n)
\end{eqnarray}
In the presence of a large Mott-Hubbard gap, all these three symmetry breaking
terms are of the order of $U$, {\it i.e.} $\chi_c^{-1} \sim g \sim \mu_c \sim U$.
Therefore, this Lagrangian contains high energy degrees of freedom of the order
of $U$. However, as already observed in \cite{so5}, at $\mu=\mu_c=\sqrt{g/\chi_c}$, 
their effects cancel completely in the time-independent part of the 
Lagrangian, and the static potential is $SO(5)$ symmetric just as in the original 
unprojected model. We also observe that near the AF/SC transition point 
where $\mu\sim \mu_c$, the first order time derivative term is of the order of one.
Furthermore, in the spirit of the low frequency and wave vector expansion, we
only need to retain the first order time derivative term in the charge sector
and can drop the second term in the above Lagrangian. Combining all these considerations,
we obtain the following low energy effective Lagrangian near the AF/SC transition
region, which is free of any parameters of the order of $U$:
\begin{equation}
{\cal L} = \frac{\chi_s}{2} (\partial_t n_\alpha)^2
+ (n_5\partial_t n_1 - n_1\partial_t n_5) 
-\frac{\rho}{2} (\partial_k n_a)^2 - V(n)
\label{projected_L}
\end{equation}
This is exactly the Lagrangian counterpart of the $SO(5)$ projection
procedure discussed 
previously in the Hamiltonian language. Dropping the second order time 
derivative terms removes half of the (high energy) degrees of freedom, 
and redefines the canonical conjugacy of the dynamical variables. In particular,
the conjugate variable of $n_1$ is nothing but $n_5$ itself, since
$p_1=\delta {\cal L}/\delta \partial_t n_1 = 2 n_5$. Standard quantization procedure
requires the canonical commutation relation $[n_1,p_1]=i$, 
which in this case just reproduces equation (\ref{non-commute}). This confirms
the fact that the $SO(5)$ projection does not change the form of the interaction
potential, only the commutation relation between $n_1$ and $n_5$.

It is easy to see that the low energy effective Lagrangian 
(\ref{projected_L}) produces exactly the same long wave length
collective mode spectrum as
the projected $SO(5)$ Hamiltonian (\ref{Hamiltonian}) defined
on a lattice. To facilitate the comparison, we take the 
$SO(5)$ potential to be
\begin{equation}
V(n) = -\frac{\delta}{2} \sum_a n_a^2 + \frac{W}{4} n_a^4 \ \ , \ \ \delta>0   
\end{equation}
Assuming broken symmetry in the $n_1$ and $n_2$ directions, we find that
the $n_3$ and $n_4$ modes always decouple, and they have a linear
spin wave dispersion relation with $v_s=\sqrt{\rho/\chi_s}$.  
The Euler-Lagrangian equations of motion gives the following 
dispersion relation for the $n_1$ and $n_2$ modes:
\begin{equation}
\omega^2 = \left\{ \begin{array} {l}
    \frac{1}{4}\rho^2k^4 \\
    \frac{\rho}{\chi_s}k^2 +\frac{2\delta}{\chi_s}
  \end{array} \right. 
\end{equation}
for the AF state with $\langle n_1\rangle =0, \langle n_2\rangle \neq 0$, 
\begin{equation}
\omega^2 = \left\{ \begin{array} {l}
    \frac{1}{2}\rho\delta k^2 \\
    \frac{\rho}{\chi_s} k^2
  \end{array} \right. 
\end{equation}
for the SC state with $\langle n_1\rangle \neq 0, \langle n_2\rangle =0$, and 
\begin{equation}
\omega^2 = \left\{ \begin{array} {l}
   \frac{1}{4}(1+\sin^2\alpha/\cos^2\alpha)\rho^2k^4 \\
   \frac{2\delta \cos^2\alpha}{\chi_s} +(\frac{1}{2}\delta \rho \sin^2\alpha+ \frac{\rho}{\chi_s}) k^2
  \end{array} \right. 
\end{equation}
for the mixed state with $\langle n_1^2\rangle  = \frac{\delta}{W} \sin^2\alpha$ and 
$\langle n_2^2\rangle  = \frac{\delta}{W} \cos^2\alpha$.
These dispersion relations agree exactly with the lattice model results at the
projected $SO(5)$ symmetric point if we make the following identification
\begin{equation}
 \rho = 2J \ \ , \ \
 \chi_s = \frac{1}{2J} \ \ , \ \
\delta = 2(4J -\Delta)
\end{equation}
The effective Lagrangian can be easily used to discuss effects of $SO(5)$ symmetry
breaking. The simplest form of symmetry breaking is increasing the chemical potential
beyond the critical value $\mu_c$, so that a pure SC state is realized. The
chemical potential enters the effective Lagrangian through the gauge coupling
in the time direction via the following substitution
\begin{equation}
\partial_t n_1 \Rightarrow \partial_t n_1 + \delta \mu n_5 \ \ , \ \ 
\partial_t n_5 \Rightarrow \partial_t n_5 - \delta \mu n_5 \ \ , \ \ 
\end{equation}
where $\delta\mu\equiv \mu-\mu_c$ is the deviation of the chemical potential
away from the critical value. In this case, the spin triplet excitations 
acquire a finite mass gap, with the following dispersion relation:
\begin{equation}
\omega^2(k) = \frac{\rho k^2}{\chi_s} + \frac{4(\mu-\mu_c)}{\chi_s} 
\end{equation}
and the mass gap increases with increasing doping in the SC state.

\begin{figure*}[h]
\centerline{\epsfysize=6.0cm 
\epsfbox{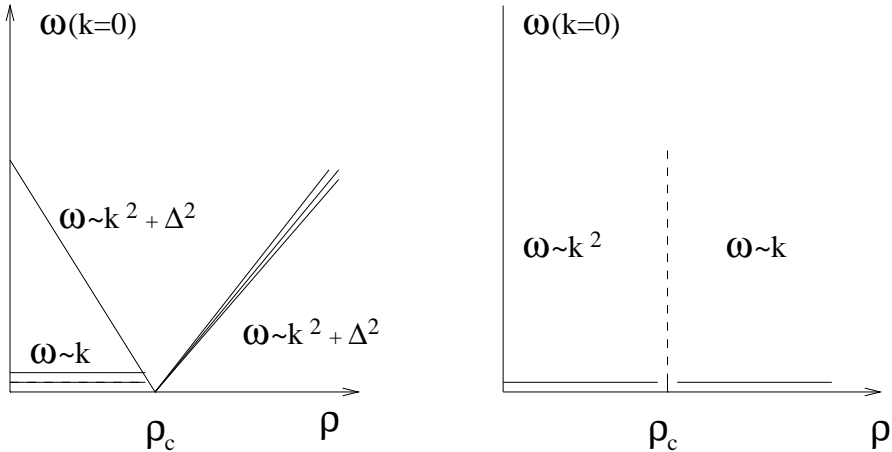}
}
\caption{Evolution of the collective mode spectra as a function of density
in the projected $SO(5)$ model. Fig. 5a shows the gap towards spin excitations.
Charge excitations are gapless for the entire region of density, however,
the dispersion relation changes from $\omega\sim k^2$ to $\omega\sim k$
at $\rho_c$, as indicated in Fig. 5b.}
\label{fig5}
\end{figure*}

We summarize the behavior of the collective modes obtained in the
previous two sections in Fig. (\ref{fig5}).

\begin{figure*}[h]
\centerline{\epsfysize=8.0cm 
\epsfbox{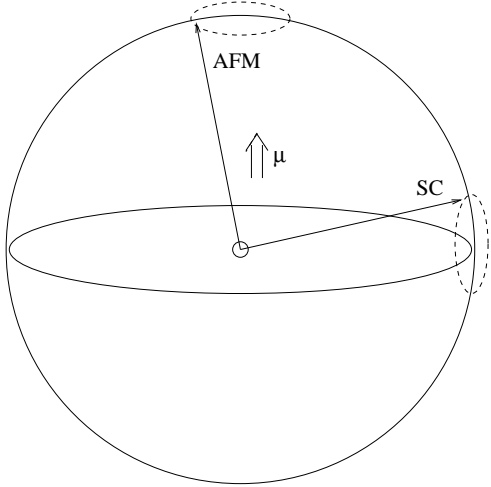}
}
\caption{Pictorial representation of a chiral $SO(5)$ sphere. }
\label{fig6}
\end{figure*}

We see that while there are significant modifications of the collective 
mode spectra in the density regime $0<\rho<\rho_c$ from the unprojected
$SO(5)$ symmetry, the spectra beyond $\rho_c$ is essentially identical to
the behavior expected from the unprojected $SO(5)$ symmetry. This should
be expected from our general considerations about the Gutzwiller projection
without much detailed calculations. We argued that the only effect of the
Gutzwiller projection is to change the quantum commutation relation
between the SC components of the superspin $n_1$ and $n_5$. However, for
$\rho>\rho_c$, the system is in a pure SC phase where these components
acquire classical expectation values. In this case, the modification of 
the quantum commutation relation does not have any significant effect. 
This argument can also be illustrated by a simple picture of a {\it chiral}
$SO(5)$ sphere, as depicted in Fig. (\ref{fig6}).

In this picture, the north and
south poles represent the three AF directions, and the equatorial plane represent
the SC directions. The sphere is perfectly $SO(5)$ symmetric. However, the 
chemical potential along the pole direction acts like a fictitious magnetic
field which restricts the sense of the rotation in the SC plane.
Small oscillations
of a vector pointing close to the north pole enclose the fictitious magnetic
flux, and can only execute chiral rotations. This amounts to the projection of 
the particle-pair states at half-filling. On the other hand, small oscillations
of a vector pointing anywhere along the equator does not enclose the fictitious
magnetic flux, and their dynamics is therefore unaffected by the projection.
Dynamics of a vector pointing anywhere between the north pole and the equator
is also partially affected by the projection, but the symmetry of the static
potential bears a unique signature.

\section{Conclusion}
\label{sec:VII}

The main purpose of this paper is to introduce the concept of
projected $SO(5)$ models and discuss the properties of this
model in connection with high $T_c$ superconductivity.
The projected $SO(5)$ model describes the low energy and
long distance bosonic degrees of the freedom near the
AF/SC transition. We showed that the Gutzwiller projection
can be implemented analytically on every site in the
$SO(5)$ theory. In the presence of a infinite Mott-Hubbard
gap, we show that static properties of the model can remain
$SO(5)$ symmetric, while the modification of the dynamics
can be completely cast into a non-trivial commutation
relation between the two SC components of the $SO(5)$
superspin. Unlike the unprojected $SO(5)$ models which 
can only have the full dynamic $SO(5)$ symmetry at half-filling,
the projected $SO(5)$ model can have static $SO(5)$ symmetry
at a critical value of the chemical potential $\mu_c$ and for a finite
range of doping $0\le\rho\le\rho_c$. At $\rho=0$, the system has
a AF ground state and zero compressibility. In the intermediate regime
$0<\rho<\rho_c$, the system has mixed
AF and SC order and infinite compressibility. For $\rho>\rho_c$,
the system has a pure SC ground state, the SC order parameter
rises to a maximal value before it decreases with doping.
At the projected $SO(5)$ symmetric point, we can understand 
precisely the evolution of the collective modes. On the
AF side, we have two gapless spin wave modes and a gapless
charge mode describing the gapless fluctuation from AF to SC.
In the intermediate density regime $0<\rho<\rho_c$, the physical
properties of the spin waves remain unchanged, while the massive
spin amplitude mode gradually decreases its energy and merges
with the two spin wave modes at $\rho=\rho_c$. The charge mode
in the intermediate density regime is gapless, but has quadratic
dispersion relation, which is a unique signature of the projected
$SO(5)$ symmetry. For $\rho>\rho_c$, the charge mode is gapless
with linear dispersion relation, and the $\pi$ triplet spin
mode becomes massive, and gradually increases its energy with
increasing doping. In this regime, the behavior of the collective
modes are identical to the unprojected $SO(5)$ model.

This very simple model can form the basis to understand many
novel and puzzling properties of the high $T_c$ superconductors
in a unified framework. It points out a route from AF to SC through
a gradual rotation of the superspin angle. At the projected $SO(5)$
symmetry point the mean field energy is independent of the superspin angle, and
therefore it offers an explanation
of the absence of the chemical potential shift in the underdoped
regime without global phase separation. It predicts a phase diagram
which is qualitatively consistent with the observed phase diagram
in the high $T_c$ materials. In the underdoped regime of the phase
diagram, the system have large AF and SC fluctuations, and these
fluctuations can be responsible for the pseudogap physics observed
in these materials.
 
There are many possible directions to carry out this line of research
in the future. The most important issue is to understand the precise
nature of the intermediate state in the regime $0<\rho<\rho_c$. 
Since the system has infinite compressibility in this regime, 
different small perturbation may select different ground states.
Such perturbing effects might include quantum fluctuations and 
longer ranged interactions. In particular, we would like to 
investigate the possibility that these perturbations might lead
to the formation of incommensurate order or stripes. 

In this work, we have discussed extensively the collective 
fluctuations in the long wave length limit. Due to the definitions
of our effective lattice model, the $k\rightarrow 0$ limit 
corresponds to the $k\rightarrow 0$ limit in the SC correlation
functions and the $k\rightarrow (\pi,\pi)$ limit of the AF
spin correlation functions. Within the $SO(5)$ theory, the
$\pi$ resonance in the SC state is viewed as the $SO(5)$ symmetry
partner of the $k\rightarrow 0$ Goldstone mode of the SC
phase fluctuation. While the commensurate neutron resonance 
mode is observed in both YBCO and BISCO superconductors, 
all high $T_c$ systems also have incommensurate spin 
fluctuations. How can these features be explained within the
current theoretical model?

The fact that LSCO and YBCO have very different Fermi surface 
shapes and yet have similar incommensurate magnetic peaks 
strongly suggests that the incommensurate peaks are not 
sensitive to Fermi surface effects and should be explainable
within an effective bosonic model. 
Let us recall that the collective mode of a superfluid boson
system consists of linearly dispersing phonon branch and
another roton branch with a minimum located at the inverse 
inter-particle spacing. So far, we have only studied the 
phonon branch of the charged bosons. By analogy, the roton
branch should also exist, with a wave vector determined by 
the density of the charged bosons or doping. Within the
$SO(5)$ theory, while the commensurate neutron resonance 
can be viewed as the $SO(5)$ partner of the SC phase mode,
the incommensurate magnetic peaks can be viewed as the
$SO(5)$ partner of the roton minimum of the charged bosons.
A detailed quantitative analysis of this picture will be carried
out in the future.

However, while the ground state in the doping range $0<\rho<\rho_c$
may depend sensitively on small perturbation effects,
at finite temperature, these perturbation effects
should be small and the system should display more universal
properties. We have shown that the projected $SO(5)$ symmetry 
should be valid for the entire doping range $0<\rho<\rho_c$,
and we shall quantitatively study the manifestation of this 
symmetry at finite temperature, and see if the projected 
$SO(5)$ symmetry can give a universal explanation
of the pseudogap physics. 

{\it Note Added:} After completing this work, we received a very interesting
paper by Coen van Duin\cite{coen}, in which he also observed the
``remnant $SO(5)$ behavior in the large $U$ limit".

\section*{Acknowledgments}
We would like to acknowledge useful discussions with D. Arovas,
 J. Berlinsky, E. Demler, R. Eder, C. Kallin,
S. Kivelson and D. Scalapino.
SCZ and JPH are supported by the NSF under grant numbers DMR-9814289.
WH and EA are supported by FORSUPRA II, BMBF (05 605 WWA 6) 
and by the Deutsche Forschungsgemeinschaft (AR 324/1-1)
and (HA 1537/17-1).
AA is supported by the Israel Science Foundation. AA,
EA and WH would like to acknowledge the support and hospitality of the
Stanford Physics department, where most of this work was carried out.

\appendix
\section{Slave Boson Results}
\label{sb}

Alternatively, one can enforce
the hard-core constraint
(\ref{hard-core})
by introducing
an additional ``slave'' boson  for
each lattice site. The presence of this  boson ($e(x)$) indicates that
the lattice site $x$ is in the singlet state.
The ``less or equal'' hard-core condition (\ref{hard-core})
 is  replaced with the
equality constraint
\begin{equation}
Q(x)= \sum_{\alpha} t_{\alpha}^{\dag}(x) t_{\alpha}(x) +
t_{h}^{\dag}(x) t_{h}(x)
+ e^{\dag}(x) e(x) - q  =0 \;,
\label{constraint}
\end{equation}
with $q=1$.
Since in physical states one always has one and only one boson per
lattice sites, destruction (creation) of a boson $t_a$ ($t_a^{\dag}$) 
must always be accompanied by creation (destruction) of the empty boson
 $e^{\dag}$ ($e$). In this way, the {\it physical} operators for creating 
 (destroying) a triplet ($a=x,y,z$) or a hole pair ($a=h$)  acquire
 the form
$ t_a^{\dag} e$ ($ t_a e^{\dag}$).
The advantage of this method is that the constraint can be enforced
exactly (at least in principle) by
 introducing an
additional time-independent field $\lambda(x)$ at each lattice site,
 by adding to the Hamiltonian a term $ - \lambda(x) Q(x)$ and by
 integrating over the $\lambda(x)$ on the imaginary axis.
The whole Hamiltonian (\ref{Hamiltonian}) 
thus takes the form (apart for a constant)
\begin{eqnarray}
\label{hsb}
&& 
H_{sb} = - \sum_x \lambda(x) Q(x)
+  \Delta_s \sum_{x,\alpha} 
    t_{\alpha}^{\dag}(x) t_{\alpha}(x) +
  \tilde \Delta_c \sum_{x} 
    t_{h}^{\dag}(x) t_{h}(x) 
\\ \nonumber &&
-
  J_s/2 \sum_{<x,x'>,\alpha} 
    ( t_{\alpha}^{\dag}(x) e(x) + h.c.)
    (t_{\alpha}^{\dag}(x') e(x') + h.c.)
\\ \nonumber &&
-
  J_c/2 \sum_{<x,x'>} 
     \left( t_{h}^{\dag}(x) e(x) e^{\dag}(x')
    t_{\alpha}(x') e(x') + h.c. \right) \;.
\end{eqnarray}

In practice, one starts with a mean-field approximation and expands
the boson operators around their mean-field values as in Sec. \ref{modes}.
This expansion can be rigorously controlled by generalizing the
constraint (\ref{constraint}) to large values of $q$,
whereby one        
scales $J_{c/s} \to  J_{c/s}/s $
(cf. Refs. \cite{re.ne.83,fi.phr}).
Physically, this corresponds to allow for a large number of bosons to
be present at each site and thus to have a large value for the total
spin, or more precisely for the SO(5) quantum number $s$, at each site.
The mean-field
result thus corresponds to the $q\to\infty$ limit, while the quadratic
expansion corresponds to the first $1/q$ correction.

At the mean-field level, the constraint is fulfilled exactly and indeed
one obtains the same result
and the same phase diagram
as the variational
ansatz (\ref{variational}) discussed in Sec. \ref{phase}.
By expanding the bosons quadratically
around the mean-field
one obtains the same  modes as for the soft-core Hamiltonian with a
similar dispersion.
(Here, we restrict again to the SO(5)-symmetric case).
Specifically, in
the mixed phase one obtains two spin-wave modes with dispersion
\begin{equation}
\omega(k) = \frac{4 J + \Delta }{4} \ k   \;,
\end{equation}
 one massive spin-amplitude mode
\begin{equation}
\omega(k)^2 = (16 J^2-\Delta^2) \cos^2\alpha + O(k^2) \;,
\end{equation}
and a quadratic mode
\begin{equation}
\omega(k) = \frac{4 J+\Delta}{8 \cos\alpha} k^4 \;.
\end{equation}
In the pure superconducting phase,
one has a $\pi$-triplet with dispersion
\begin{equation}
\omega(k) = \frac{4 J + \Delta }{4}  \ k 
\end{equation}
and the SC Goldstone mode with dispersion
\begin{equation}
\omega(k)^2 = \frac{16 J^2 - \Delta^2}{4}\   k^2 
\end{equation}
These results coincide with the ones of the soft-constraint
approximation
Eqs. (\ref{spin_wave}) to (\ref{mixed_modes})
in the limit of small $4 J -\Delta$, i. e. ,  for low boson
density.


\end{document}